\documentclass[draftcls,onecolumn]{IEEEtran}
\usepackage{graphicx,subfigure,epstopdf,times,amsmath}
\usepackage{amssymb}
\usepackage{enumerate}
\usepackage[T1]{fontenc}
\usepackage{float}
\usepackage{multirow}
\usepackage{color}

\title{FPGA Hardware Acceleration of Monte Carlo Simulations for the Ising Model}
\begin{document}
\author{Francisco~Ortega-Zamorano, Marcelo A. Montemurro, Sergio A. Cannas, Jos\'e~M.~Jerez, and Leonardo~Franco. \thanks{F. Ortega-Zamorano, J.M. Jerez and L. Franco  are with the Departamento de Lenguajes y Ciencias de la Computaci\'on, Universidad de M\'alaga, Campus de Teatinos S/N, 29071, M\'alaga, Spain. S.A. Cannas belongs to Facultad de Matem\'atica, Astronom\'{\i}a y F\'{\i}sica (IFEG-CONICET), Universidad Nacional de C\'ordoba,  C\'ordoba, Argentina. M. A. Montemurro is at the Faculty of Life Sciences, University of Manchester, Manchester, UK. Correspondence should be addressed to Francisco Zamorano (fozamorano@gmail.com)}}

\maketitle

\begin{abstract}
A two-dimensional Ising model with nearest-neighbors ferromagnetic interactions is implemented in a Field Programmable Gate Array (FPGA) board.
Extensive Monte Carlo simulations were carried out using an efficient hardware representation of individual spins and a combined global-local LFSR random number generator. Consistent results regarding the descriptive properties of magnetic systems, like energy, magnetization and susceptibility are obtained while a speed-up factor of approximately 6 times is achieved in comparison to previous FPGA-based published works and almost $10^4$ times in comparison to a standard CPU simulation. A detailed description of the logic design used is given together with a careful analysis of the quality of the random number generator used. The obtained results confirm the potential of FPGAs for analyzing the statistical mechanics of magnetic systems.
\end{abstract}

\section{Introduction}
In recent years several advances in microelectronics have permitted the utilization of new devices for carrying out parallel numerical simulations in order to speed up the computational times involved. Among the most used hardware devices we can mention multi-core processors \cite{Vajda2011}, GPU cards \cite{Suchard2010} and FPGA boards \cite{Sano2014}. As it is usually the case, there is no one system better than other for all situations, as the answer is very much dependent on the problem under analysis together with  the circumstances of the developers in relationship to budget, programming expertise, development time, etc. \cite{Weber2011}
In this work a FPGA based implementation of the two-dimensional ferromagnetic Ising model is carried out using Monte Carlo simulations.
Field Programmable Gate Arrays (FPGA) are reconfigurable hardware devices that can be reprogrammed to implement different combinational and sequential logic created with the aim of prototyping digital circuits as they offer flexibility and speed. In recent years advances in technology have permitted to construct FPGAs with considerable large amounts of processing power and memory storage, and as a consequence they have been applied in an ever increasing range of domain, like telecommunications, robotics, pattern recognition tasks, and infrastructure monitoring, among others \cite{Song2011}, \cite{Monmasson2011}, \cite{Ganegedara2014}.

The Ising model is a paradigm of the statistical physics approach to the study of finite temperature equilibrium properties of  many body systems, by reducing complex interactions to their minimal expression (for a short review see Ref.\cite{Binder2001} and references therein). The model assigns a set of binary variables, called spins, to every site of a regular grid or lattice. Every spin represents the component of the local magnetic moment in a crystalline solid, respect to the direction of an external magnetic field. The model is completed by defining an energy function that depends on the values of all spins in the lattice. Statistical Mechanics provides then a recipe to derive from the energy function the equilibrium probability distribution of the microscopic configurations of the system (i.e., any combination of spin values), from which macroscopic properties---like, the total magnetization or mean energy---are obtained by averaging the appropriate variables.

Despite its relative simplicity, Ising-type models and their generalizations (e.g., Potts model, see \cite{Ferrero1} and references therein) are extensively used to analyze properties of a large variety of systems exhibiting cooperative phenomena, ranging from simple ferromagnetism to complex disordered materials (e.g., spin glasses) \cite{Wolf2000}. Moreover, being originally restricted to the realm of solid state physics, they have been shown to be extremely useful in other disciplines, such as soft condensed matter (e.g. soap bubbles and foam \cite{GlWe1992,SaGl2006}), biology (e.g., biological cells \cite{GrGl1992})  and neural networks \cite{Amit}.

The cooperative phenomena those models intend to describe are a synergistic result of the interaction among a very large (macroscopic) number of relatively simple units.  Statistical Mechanics theory seeks then to describe the asymptotic behavior of averaged macroscopic quantities---like energy or magnetization---in the limit of an infinite number of units (the so called thermodynamic limit). However, in most cases it is extremely difficult to obtain such limiting behavior analytically. Thus, the usual approach is to perform numerical simulations for an increasing number of units and then extrapolate the results to the limit of infinite number of units. This raises the question about the minimum number of units needed to obtain an accurate extrapolation to the thermodynamic limit. While for the estimation of some quantities up to $\sim 10^4$ units may be enough (and attainable in a CPU code in a few hours), others may require at least $\sim 10^6$ or more units to get reliable results (see \cite{Ferrero1} and references therein). Such calculations are only achievable in reasonable time scales with the use of parallel computing. Therefore, it is of critical importance to develop and assess massively parallel implementations of statistical physics models. In that respect, the Ising model has become one of the most common benchmarks for testing novel statistical mechanics simulation algorithms and parallel computing implementations. The most simple version (namely, that in which only nearest neighbors spins interact ferromagnetically) constitutes one of the few examples of an interacting many body system for which non trivial equilibrium properties are known exactly in two spatial dimensions \cite{Huang}. This is not true in general for systems with more complicated interactions. Hence, theoretical analysis is usually limited to approximated methods whose ultimate validity strongly relies on Monte Carlo numerical simulations.

  One of the most common ways to simulate the behavior of Ising-type models is the Metropolis-Hastings algorithm \cite{Hastings1970}, which works by generating a sequence of sample configurations of the system that converges to the equilibrium finite temperature distribution. The Metropolis-Hastings algorithm requires the use of random numbers and as such is considered a Monte Carlo type simulation \cite{Binder1995}. Monte Carlo simulations depend strongly on the generation of pseudo random numbers and for this reason an efficient simulation normally adapts the algorithms used to the hardware utilized. In particular, in FPGA boards the standard random number generators are based on Linear Feedback Shift Registers (LFSR) schemes \cite{Hurd1989}.
	
FPGA boards can be used as hardware accelerators systems in several domains of applications. For example in life sciences, several recent works have benefitted from the use of FPGA boards for speeding up Monte Carlo simulations. The use of FPGA boards to study Ising type systems is relatively recent and thus just very few works have been published so far. Among these, the approach taken by the Janus consortium is worth mentioning as they are using a cluster of FPGA boards \cite{Baty,Belleti2009}. Lin {\em et al}. have studied the two-dimensional Ising model  \cite{Lin2013}, and more recently Gilman have analyzed the 3-d Ising  model \cite{Gilman2013}.
In this work, further optimization of the parallel updating of spin blocks is achieved by combining a global 32 bit LFSR with a smaller 12-bit local LFSR to get random numbers for the individual spin updates. Based on that strategy, we developed a highly efficient FPGA implementation of the Metropolis-Hastings algorithm for Ising type models, which was checked against the exact results for the two dimensional ferromagnetic nearest-neighbor models. The present implementation shows a considerable performance improvement with respect to previous ones.

\section{The two dimensional Ising model}
The Ising model was originally devised to represent a ferromagnetic solid. It assigns a binary variable $S_i=\pm 1$, called spin, to each site of a regular lattice in $d$ dimensions, where $i$ labels the site. $S_i$ represents the component of the magnetic moment at site $i$ respect to the direction of an external magnetic field of intensity $B$.
In this work we have analyzed the 2-dimensional Ising model with  interactions between  nearest neighboring spins, in a square lattice with $N=L\times L$ sites,
For this case, the energy of the system (Hamiltonian) is defined as:

\begin{equation}
H = -J \;  \sum_{<i,j>} S_i S_j  - B \; \sum_{i=1}^N  S_i,
\label{EcuHamilton}
\end{equation}

\noindent where $<i,j>$ denotes a sum over all pairs of nearest neighbor sites of the lattice and $J>0$ is the ferromagnetic interaction constant or also called exchange constant. In the absence of a magnetic field $B=0$, the energy is minimized in the ferromagnetic state, i.e., when all spins take the same value. In this work we considered $J=1$  and $B=0$, and used periodic boundary conditions.

The  spin dynamics is governed by the Metropolis-Hastings algorithm, which essentially consists in the update of the present state of a given spin according to the change in the energy $\Delta E$ produced by the flipping of the spin. The spin value is changed if its flip reduces the energy, i.e. if $\Delta E<0$. If  $\Delta E>0$ the spin flip can be accepted, provided that a generated random number $r$ is smaller than the Boltzmann factor

\begin{equation}\label{EcuComparison}
    r < e^{-\beta\Delta E}
\end{equation}

\noindent where $\beta=1/k_B T$, $T$ being the temperature and $k_B$ is the Boltzmann constant (in this work we use $k_B=1$). In a sequential implementation the algorithm proceeds by picking spins one by one according to some protocol (usually randomly) until all spins (on the average in the case of random picking) have been updated (an update means a single trial; the spin can remain in its previous value after the update). Once all spins have been updated, this is called a Monte Carlo Step (MCS) and constitutes the basic iteration time unit.

Parallel implementations of the Metropolis algorithm in the square lattice make use of the nearest-neighbor interactions as follows. Let $S_0$ the spin to be updated. Then it is easy to see that the change in energy (\ref{EcuHamilton})  (with $B=0$) produced by a single spin flip $S_0 \to -S_0$ is given by $\Delta E = 2\epsilon$, where

\begin{equation}
\epsilon = (S_1+S_2+S_3+S_4)\cdot S_0
\label{EcuEnergy}
\end{equation}

\noindent and $S_1,\ldots,S_4$ denote the four nearest neighbors spins of $S_0$. The square lattice can be divided into two square sublattices with a checkerboard structure, in such a way that the nearest neighbors of any spin in a given sublattice belongs to the other (see section \label{Implementation} for details). Hence, all spins in a sublattice can be updated simultaneously without risk of concurrency issues. A MCS is then completed by one updating of the two sublattices.

In a typical simulation a sequence of random spins configurations is generated from some initial one, by successive application of the above algorithm. For a long enough sequence (measured in MCS) it can be shown that the probability distribution for spins configurations becomes stationary \cite{Binder1995}. Further application of the algorithm provides a sample set of configurations over which averages of quantities of interest can be calculated. A typical quantity is the average magnetization per spin:

\begin{equation} \label{EquMag}
m = \langle M \rangle/N,
\end{equation}

\noindent where $M=\sum_i S_i$ and $\langle \cdots \rangle$ stands for an average over a single equilibrated MC sequence of spins configurations. Another quantity of interest is the zero field magnetic susceptibility, which characterizes the linear response of the system to an externally applied magnetic field. The magnetic susceptibility can be computed form the fluctuations in the magnetization as follows:

\begin{equation}\label{EquSus}
  \chi = \frac{1}{k_B T N} \left(\langle M^2 \rangle -\langle M \rangle^2 \right) \,.
\end{equation}

\noindent In the thermodynamic limit (infinite lattice size) when $B=0$ and $d \leq2$, this model undergoes a second order phase transition at a very well defined critical temperature $T_c$, namely, the magnetization becomes zero for $t \leq T_c$ and different from zero for $T<T_c$. Also the susceptibility diverges at $T=T_c$ as $\chi \sim |T-T_c|^{\gamma}$, where $\gamma>0$ is a critical exponent that depends only on the dimensionality $d$ \cite{Reichl}.

\section{FPGA}
FPGAs \cite{FPGADesign} are reprogrammable silicon chips, using pre-built logic blocks and programmable routing resources. They can be configured to implement custom hardware functionality, and in this sense, FPGAs are completely reconfigurable and can almost instantly change its behavior by recompiling a new circuitry configuration.

The board used for the current implementation is the Virtex-5 OpenSPARC Evaluation Platform (ML509). This device includes a Xilinx Virtex-5 XC5VLX110T FPGA that provides different connector devices: 2 USBports, 2 PS/2 ports, RJ-45  and RS-232 connectors, 2 Audio Inputs, 2 Audio Outputs, Video Input, Video Output, Single-Ended and Differential I/O Expansion.
Table \ref{tabledeviced} shows some characteristics of the Virtex-5 XC5VLX110T FPGA, indicating its main logic resources. 

\begin{table}[h]
\caption{Main specifications of the Virtex-5 XC5VLX110T FPGA related to its available slice logic.}
\label{tabledeviced}
\centering
{\begin{tabular}{ | c | c | c | c| c | }
\hline
\multirow{2}{*}{Device} & Slice 	 & Slice & Bonded & Block \\
		 & Registers & LUTs  & IOBs 	& RAM\\\hline
Virtex-5  & \multirow{2}{*}{69,120}	 &\multirow{2}{*}{69,120} &\multirow{2}{*}{34}	  & \multirow{2}{*}{148}\\
\scriptsize XC5VLX110T \normalsize &  &  &  & \\
\hline
\end{tabular}}{}
\end{table}



All computations have been performed using fixed point arithmetic, which is the standard way to work with FPGA boards.
Even if floating point operations can be codified  efficiently in FPGA boards without significant additional resources  \cite{Savich2007,Jovanovic2012}, they tend to be less efficient than fixed point arithmetic, as it is also the case for most digital circuits.

\section{Implementation}
\label{Implementation}
The FPGA implementation parallelizes the spin updates in order to compute faster the system evolution. The actual value of the spins $\pm 1$ are stored as Boolean values using memory blocks while the spin dynamics is implemented using groups of LUTs. For larger lattice sizes, the number of LUTs available is not enough to simulate the dynamics of the whole lattice so the process is carried out in a number of steps in which a group of spin rows is updated.

\begin{figure*}[t]
\centering
\includegraphics[width=0.85\textwidth]{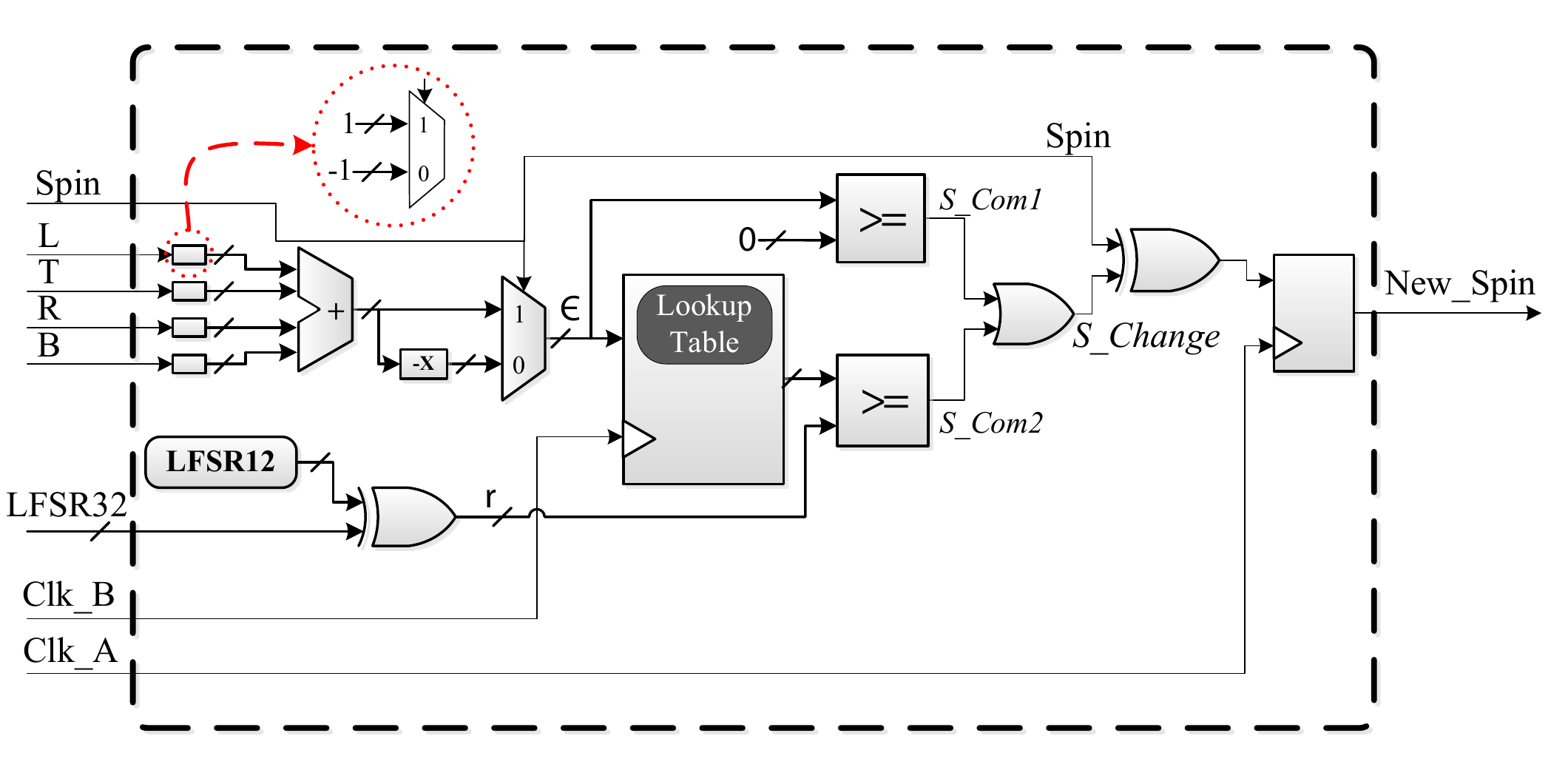}
\caption{Logic circuit representation used for the dynamic simulation of an Ising spin (see text for details).}
\label{FigIsing}
\end{figure*}

\begin{figure}[t]
\centering
\includegraphics[width=0.85\textwidth]{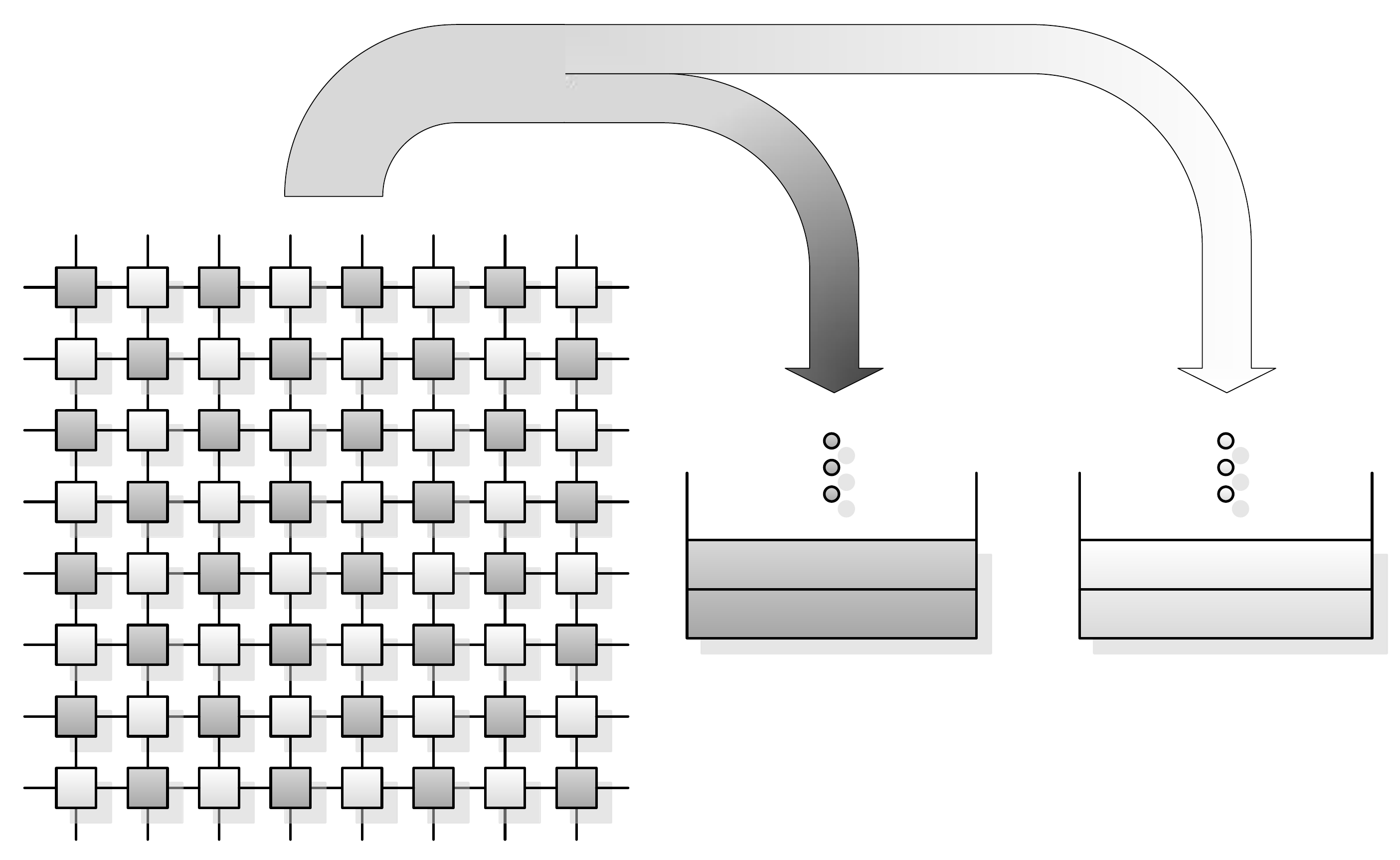}
\caption{Schematic representation of an 8x8 lattice showing the checkerboard division into two sub-lattices and the use of two memory modules to store spin values of each of the sublattices.}
\label{FigSys1}
\end{figure}

Figure \ref{FigIsing} shows the hardware implementation of a spin. On the left side of the figure, six different inputs and two clock signals are shown. From top to bottom, we see the first input (``Spin'') that gets the spin value from the register. The following  four inputs correspond to the adjacent nearest-neighbor spins of the one under consideration, and they are indicated by `L'  (left), `T' (top), `R' (right) and `B'  (bottom). The current spin value is taken from the grey block of registers, while the four adjacent spin values are taken from the white one (See Fig. \ref{FigSys1}). The input indicated by LFSR32 is a generic (pseudo) random signal that will be combined with a local LFSR through a XOR gate for obtaining a local random value (see details of this process in Section \ref{SecLFSR}). Two clock signals are used, one corresponding to the system clock signal (Clk\_A), and another with double frequency (Clk\_B) obtained using a PLL in order to read the values of the Lookup table. The frequency of Clk\_B is twice that of Clk\_A to allow the whole updating process to be completed in one system clock cycle.

The spin updating process starts by computing the summation of the values of the four adjacent spins (this process is indicated in the left top corner of the Fig. \ref{FigIsing}), noting that first the representation of these four spins is changed from the Boolean one used in the register $(0,1)$ to a binary one ($\pm 1$)  used in the  dynamics of the system (The representation modification is indicated  within a dotted circle at the top left of the figure).
The process continues with the calculation of the local function $\epsilon$ (see Eq. \ref{EcuEnergy}), that is computed multiplexing the sum value of the four adjacent spins the current spin value.

The following step consists in the computation of the exponential of   $-\beta\Delta E$, which is done using a lookup table that stores every possible values of the function $e^{-\beta\Delta E}$.

Thus, the number of entries of the look-up table is five as these are the number of possible different values of $\Delta E =(-4, -2, 0, 2, 4)$.  The values of $e^{-\beta\Delta E}$ were represented using a word length of 12. The lookup table was implemented using Configurable Logic Blocks (CLBs),  in order to save RAM resources for storing spin configurations, as only 5 LUTs are needed.

The new value of the spin can be the same as the current one or its opposite (according to the dynamics of the model). This last case can be due to a decrease on the energy associated with the spin ($\Delta E \le 0$) or in case that there is an energy increase, if the value of   $e^{-\beta\Delta E}$ is larger than a random number (indicated as $r$ in Fig. \ref{FigIsing}). These two cases are implemented through an OR gate that receives signals from two comparison gates that evaluates the procedure previously described. The output signal of the OR gate (indicated by S\_change) is then XOR with the spin value in order to obtain the updated one.  Finally, the new updated value is registered in order to synchronize the procedure.

\begin{table}[H]
\centering
\caption{\label{tableSummary}Logic utilization for the hardware implementation of a single spin}
{\begin{tabular}{ | l | c | }
\hline
Logic Utilization & Used\\\hline
\hline
\# Slice Registers			& 17\\
\# LUTs						& 30\\
\hline
\end{tabular}}{}
\end{table}

The updating procedure just described requires the logic resources indicated in Table \ref{tableSummary}, noting that the whole process can be executed in one clock cycle with a maximum frequency of 316.156 MHz. Nevertheless, the system frequency has been set to 300 MHZ, as this is the maximum frequency that can be obtained using a Phase-Locked Loop (PLL) in order to optimize the implementation.

The values shown in Table \ref{tableSummary} determine the maximum number of spins that can be updated in one cycle according to the FPGA board specifications, number that can be obtained from the following equation:

\begin{equation}
\mbox{Max\_Spin} \le \frac{ \mbox{Available LUTS}} {\mbox{LUTs per spin}}.
\label{EqMax}
\end{equation}

\begin{figure}[t]
\centering
\includegraphics[width=0.85\textwidth]{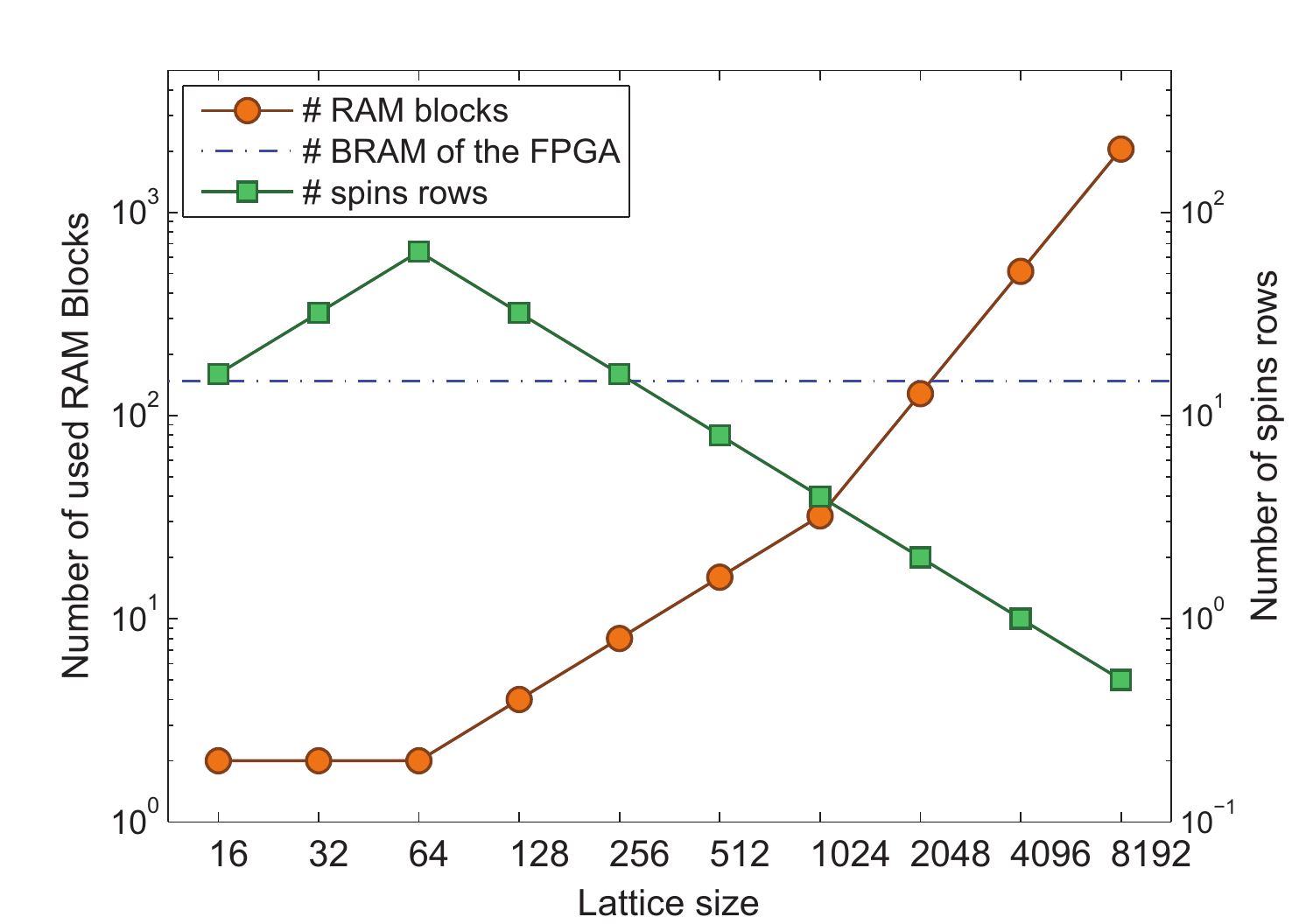}
\caption{RAM blocks resource utilization and number of spin rows that can be updated simultaneously as a function of the lattice size.}
\label{FigMemoria}
\end{figure}

 In the present case, the used board contains 69120 LUTs, permitting a simultaneous maximum implementation of $\lfloor \frac{69120}{30}  \rfloor_{2^m}=2048$ spin sites, value that is obtained from Eq. \ref{EqMax}, where the obtained number should be a power of two to optimize resource utilization. The previous analysis does not take into account the whole process as in addition the storage of the lattice spin values are needed, and this introduces a memory limiting factor. To analyze both memory and LUTS requirements we need to analyze the whole updating process considering the following. The spins are located in a 2-dimensional $L \times L$ square lattice that will be  divided in two checkerboard sub-lattices, as the whole system cannot be updated simultaneously at the risk of a feedback catastrophe \cite{Vichniac1984}. Fig. \ref{FigSys1} a) shows an 8x8 spin lattice in which two sub-lattices are shown (white and grey colors are used for differentiating them), together with two memory modules that will be used for storing these sub-lattice spin values. In our case the FPGA BRAM is organized in 148 blocks of 1024 x  36-bits long. As each of the two memory modules store only half of the total spin site, the number of used RAM blocks for different lattice sizes is indicated in Fig.  \ref{FigMemoria}. It is also shown in this figure by a horizontal dashed dotted line the maximum number of BRAM of the board that sets a maximum lattice size of 2048. Nevertheless, in order to maximize the parallelism of the procedure and in order to compare with previous works, we focused out study in the 1024x1024 case.

We describe below the whole updating process that will be carried simultaneously for a group of rows, with a limiting factor given by the maximum number of spins that can be implemented. The maximum number of rows that can be updated simultaneously can be computed simply by considering the total number of spins that can be represented according to the number of LUTs of the board ($2048$ in our case, see the analysis described above in relationship to Eq. \ref{EqMax}) divided by half of the lattice size:

\begin{equation}
\mbox { \# rows} = \frac{ \mbox{Max\_Spin}} {L/2}.
\label{EqMax2}
\end{equation}

The number of spin rows that can be updated simultaneously as just mentioned is shown in Fig. \ref{FigMemoria}.

\begin{figure}[t]
\centering
\includegraphics[width=0.85\textwidth]{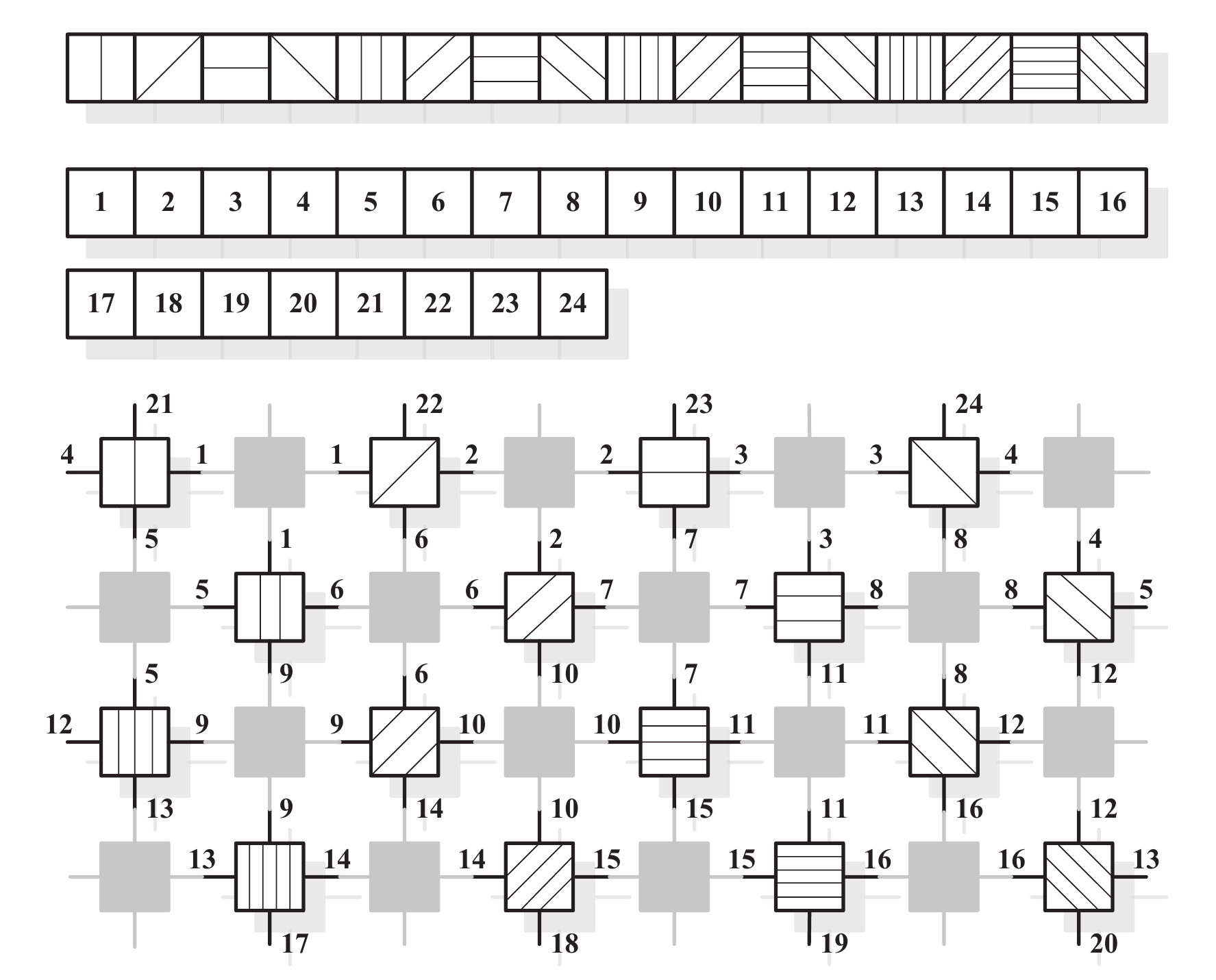}
\caption{Connectivity scheme for the update of 4 rows of an 8x8 Ising spin lattice. On top of the figure, three registers are used to store spin values corresponding to sites to be updated (first row) and to the neighboring sites (second and third rows). }
\label{FigSys2}
\end{figure}

\begin{figure}[t]
\centering
\subfigure[Step 1] {\includegraphics[width=0.25\textwidth]{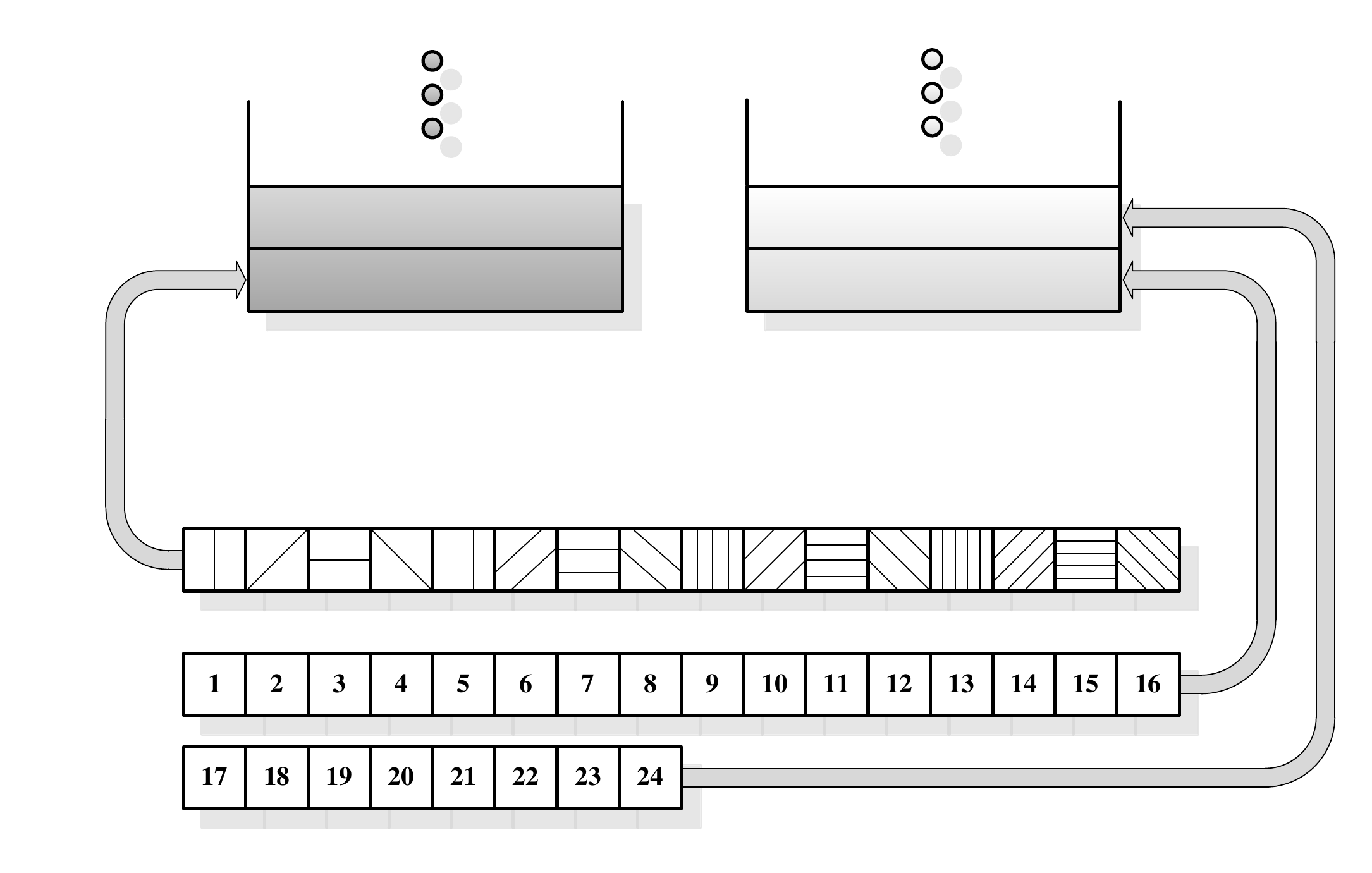}}
\subfigure[Step 2] {\includegraphics[width=0.25\textwidth]{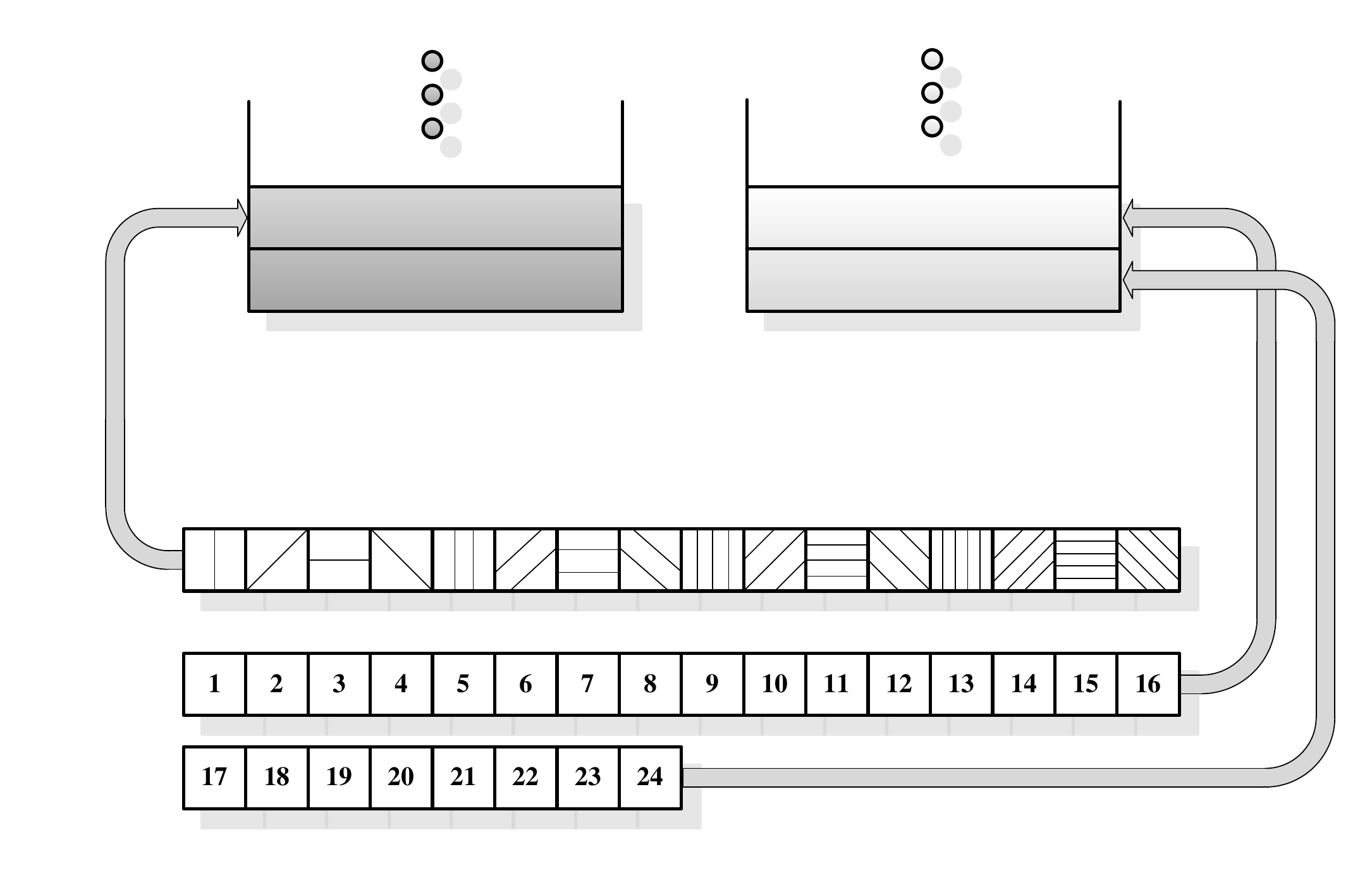}}
\subfigure[Step 3] {\includegraphics[width=0.25\textwidth]{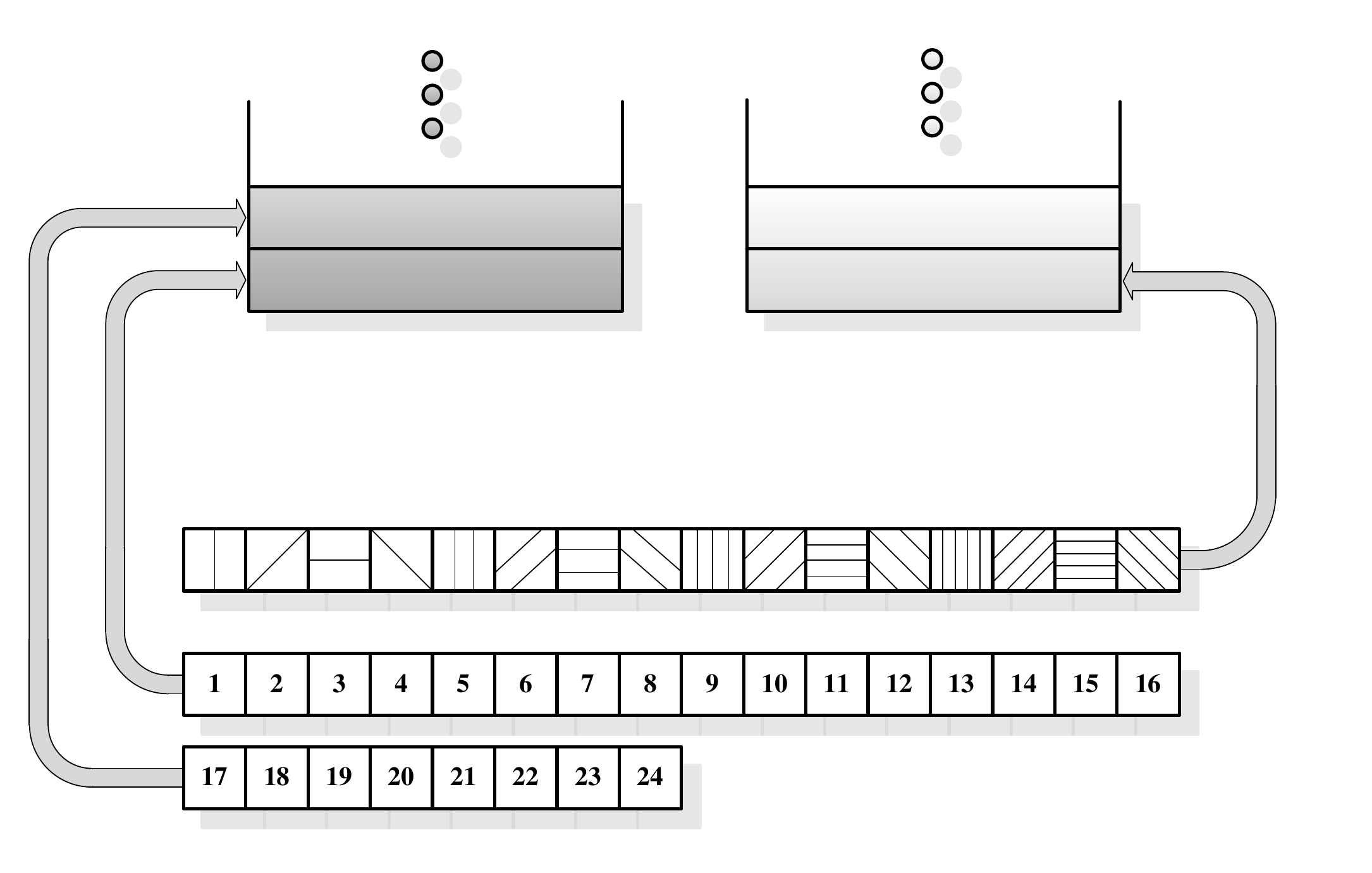}}
\subfigure[Step 4] {\includegraphics[width=0.25\textwidth]{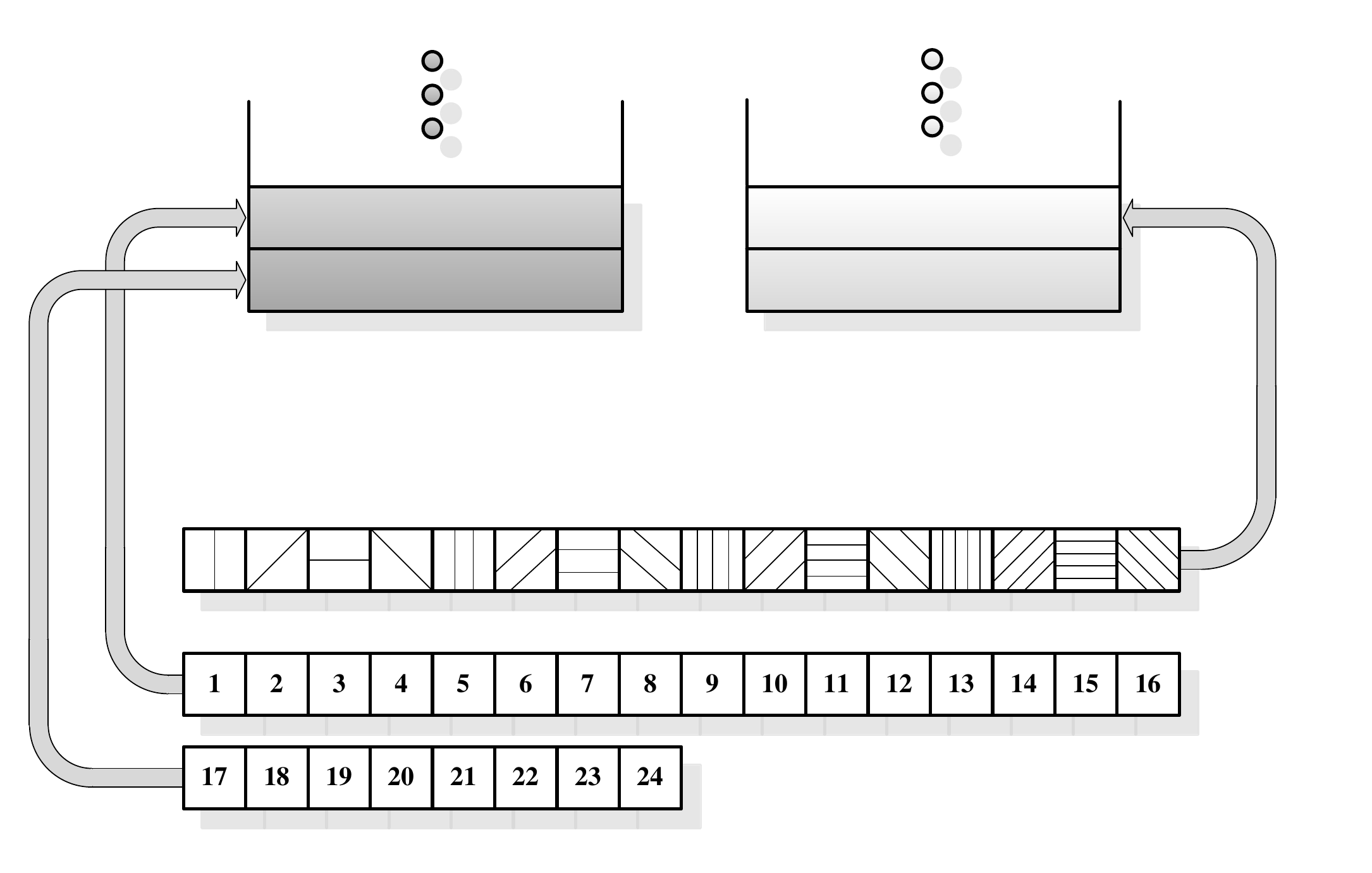}}
\caption{Schematic representation of the hardware implementation for the update procedure of a 2D 8x8 spin grid.}
\label{FigSys3}
\end{figure}

Fig. \ref{FigSys2} shows the process implemented for updating the values of one of the two sub-lattices. On top of the figure, registers for storing the values of the spin to be updated and of their neighboring ones are shown.
The first row of 16 bit register corresponds to the 16 spins that can be updated that are displayed at the bottom of the figure. The second and third rows of register are used to store the neighboring spins.
In the bottom graph the white squares are the spins to be updated and on grey color the neighboring ones, where the indicated numbers and filling pattern correspond to the codification used in the registers shown.

The whole process for updating a 8x8 grid of spins is shown in Fig. \ref{FigSys3}, where the four steps involved are shown. In the first step the first four rows of spins belonging to the grey sub-lattice (16 spins for the 8x8 size case stored in the bottom address row of the RAM) are updated (a); to then update the remaining four rows (stored in the second address row) in the second step (b). Third and fourth steps correspond to the same updating process but for the spins belonging to the white sub-lattice.

The updating process starts with an initial configuration given by random spin values and a given value of the temperature ($T$), generating blocks of memory with these initial values.
An update iteration starts by transferring the value of a row of spins to the slice registers and then to the block of LUTS that simulate the spin dynamics.

\section{LFSR random number generation}
\label{SecLFSR}
Random numbers are an important part of Monte Carlo simulations, as they should be properly generated in order to avoid
errors. As such, several authors have analyzed this topic in great depth \cite{Marsaglia2003,Matsumoto2008}. Among the different algorithms to create random numbers, Linear-Feedback Shift Register (LFSR) random number generators provide one of the most efficient alternatives to use in hardware logic implementations (FPGAs, ASICS, etc.). A LFSR works by taking a string of bits and producing the next sequence by shifting all except the rightmost bit one position to the right, setting the value of the new leftmost bit through a linear combination of the rest of the sequence. 

In a recent work Lin et al. \cite{Lin2013} have used a combination of a LFSR random number generator plus Celullar Automata (CA) in order to study a FPGA based simulation of the Ising model. Instead, in our implementation we used a novel approach based on the XOR combination of two LFSRs: a global 32-bit length and a shorter 12-bit local one. This method was chosen in order to use more efficiently the FPGA resources to allow larger numbers of spins to be updated simultaneously, and thus per unit of time, leading to a significant overall increase in computation speed  (see the detailed analysis in Table III and related text).

\begin{figure}[ht]
\centering
\includegraphics[width=0.65\textwidth]{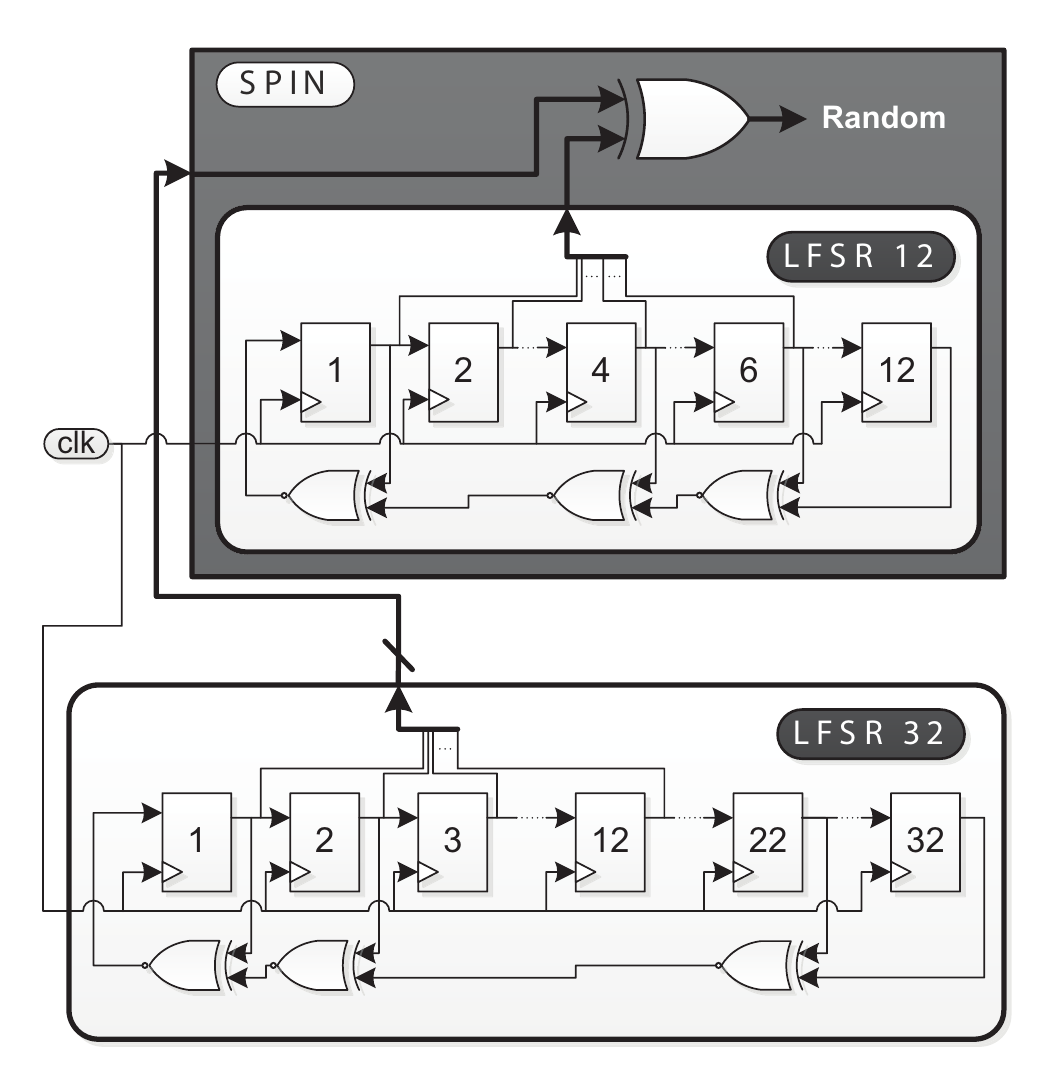}
\caption{Hardware representation of the creation of the random value in a generic spin, using a local LFSR12 and the general LFSR32.}
\label{FigLFSR}
\end{figure}

The hardware LFSR implementation consists in $M$ number of registers in series (synchronized by a unique clock signal), with $M$ being the length of the bit sequence. The output of one register is the input of the next, except for the first one, which is a linear function (normally XOR or XNOR) of some bits of the overall shift register value. 
The two modules are shown in Fig. \ref{FigLFSR}, where the top one corresponds to a part included in the implementation of every spin block. The bottom graph corresponds to a generic module that uses a 32 bits LFSR. Each spin block contains a 12 bit LFSR and the generation of pseudo random numbers for the Ising model updating is done by combining through an XOR function the 12 bits of the spin block LFSR with the first 12 bits of the sequence generated by the global LFSR32 module. 

Two different kinds of tests have been used to check for the correct implementation of the random number generation process, the first described below based on a standard suite of statistical tests (NIST) and the a second one based on visual analysis.

\subsection{\textbf{Tests for random numbers}}
The National Institute of Standards and Technology (NIST) is a measurement standards laboratory belonging to the United States Department of Commerce \cite{WebNIST}. One of its groups (the Random Number Generation Technical Working Group) has provided software containing a battery of statistical tests suitable for the evaluation of random number generators \cite{Rukhin2010}.

\begin{figure}[t]
\centering
\includegraphics[width=0.65\textwidth]{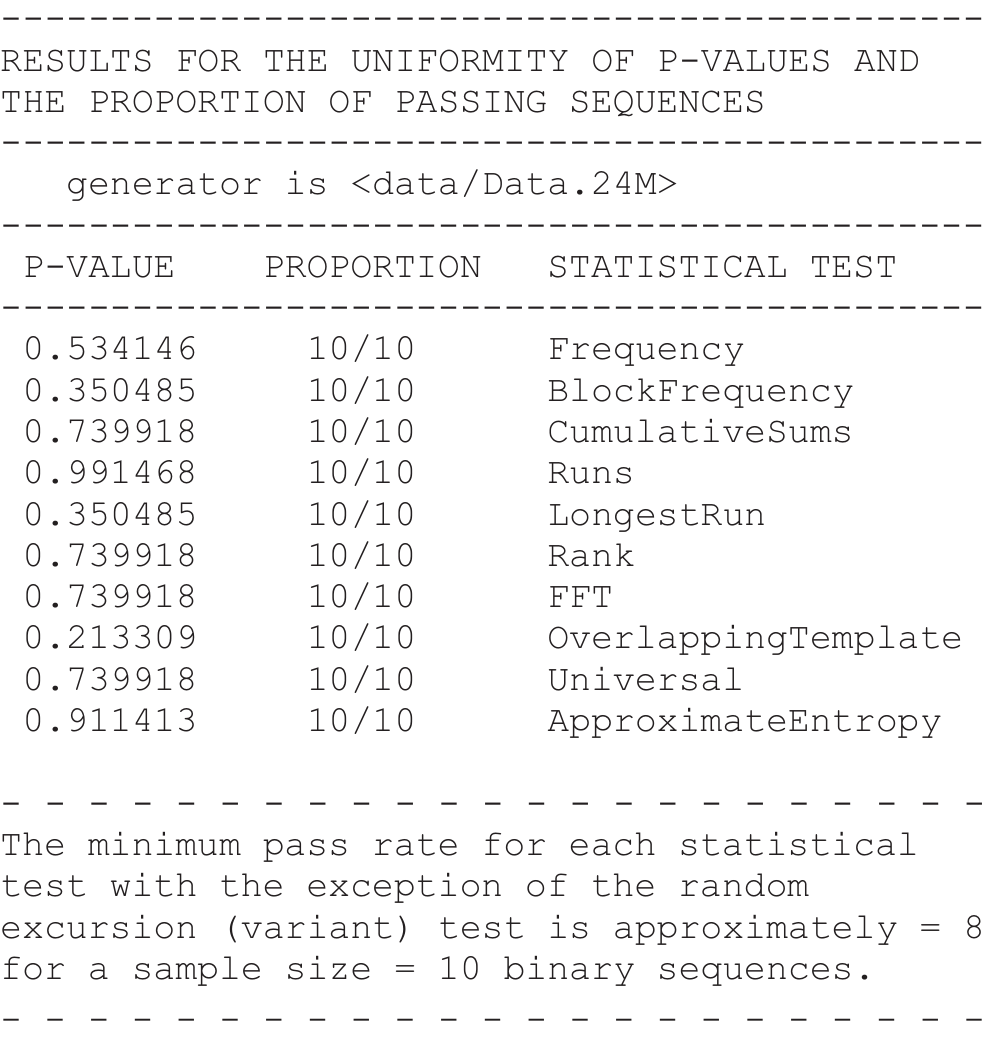}
\caption{Output screenshot of the NIST random number evaluation program that includes ten statistical tests (see text for more details).}
\label{FigResTest}
\end{figure}

We have applied the battery of 10 tests provided by NIST to  analyze the quality of the generated random numbers.
The random number generator used a combination of a 32bit global LFSR and a 12 bit local LFSR passed all the ten tests and the results are shown in Fig. \ref{FigResTest}, obtaining an average for the $p$-values of 0.6311.
As a comparison, we have also executed the tests on the  32 bit LFSR resulting in lower $p$-values (mean 0.5008),  and a failure on the Rank test.

\section{Results}

To verify the correct FPGA implementation of the Metropolis-Hastings algorithm for the 2D-Ising model, we analyzed the model's critical behavior, which is characterized by the presence of long-range spatial and temporal correlations, as well a diverging response functions\cite{Goldenfeld} (for instance, magnetic susceptibility). In this regime the system becomes extremely sensitive to external perturbations like, in particular, implementation errors in the simulation algorithm. The typical analysis of critical behavior involves the calculation of the magnetization and magnetic susceptibility as a function of the temperature for increasing system sizes.  For an infinite system size the magnetization is zero above the critical temperature $T_c$ and different from zero below it, while the susceptibility diverges at $T_c$. The respective curves for finite size systems for both quantities have to develop those singularities as the size increases, according to specific rules determined by finite size scaling \cite{Goldenfeld}. 

We started by obtaining the magnetization $m$ in the absence of an external magnetic field $B=0$ as a function of the temperature for different lattice sizes according to
 Eq. \ref{EquMag}. In order to thermalize the system 1000 MCS were executed prior to any measurement, using 1000 values for obtaining the average value for each temperature with a 100 MCS separation between consecutive measurements.

\begin{figure}[h]
\centering
\includegraphics[width=0.6\textwidth,angle=-90]{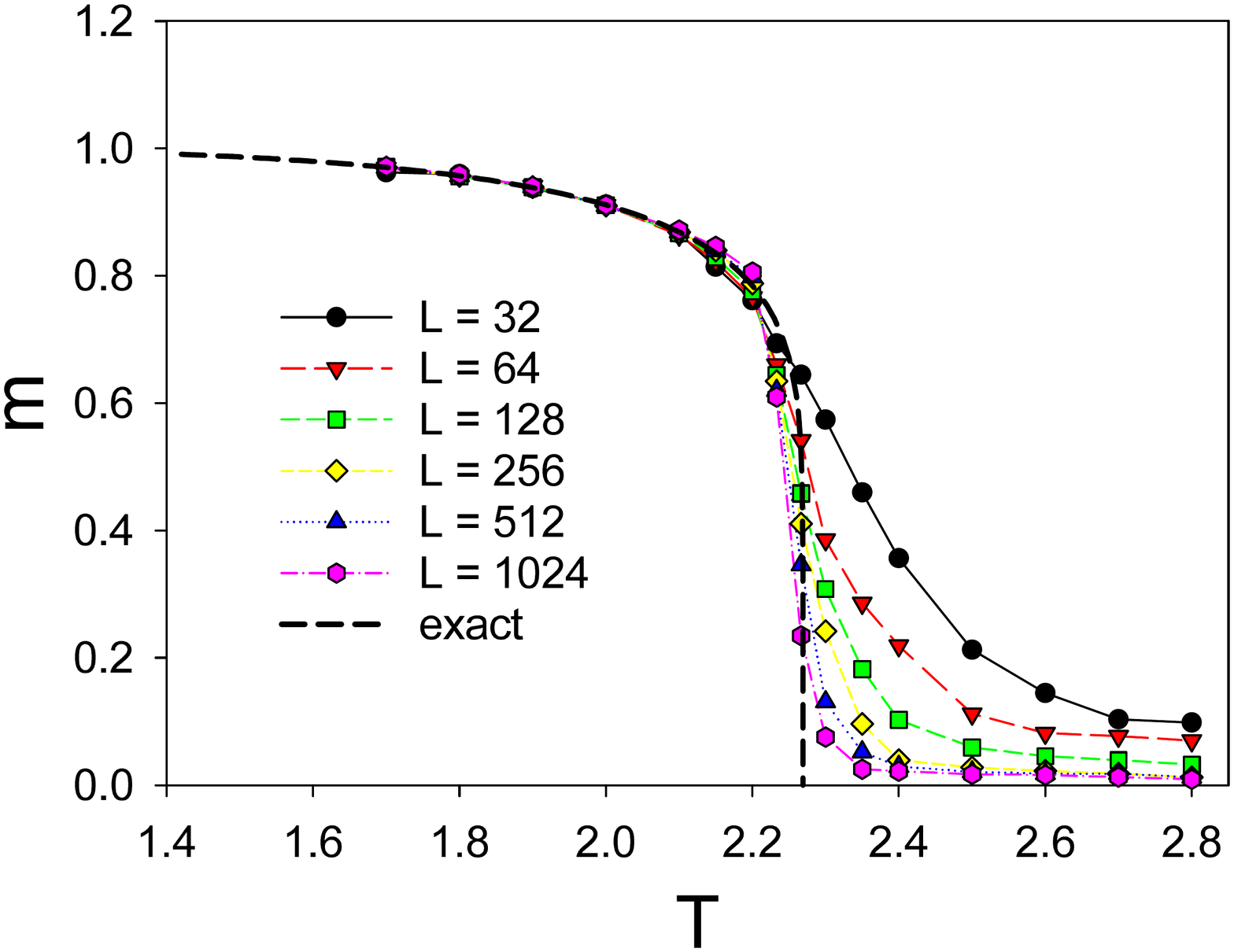}
\caption{Magnetization of a 2-D Ising model as a function of the temperature for different lattice sizes, compared with the exact result (dashed line). The continuous lines are a guide to the eye.}
\label{FigMagnet}
\end{figure}

\begin{figure}[h]
\centering
\includegraphics[width=0.5\textwidth,angle=-90]{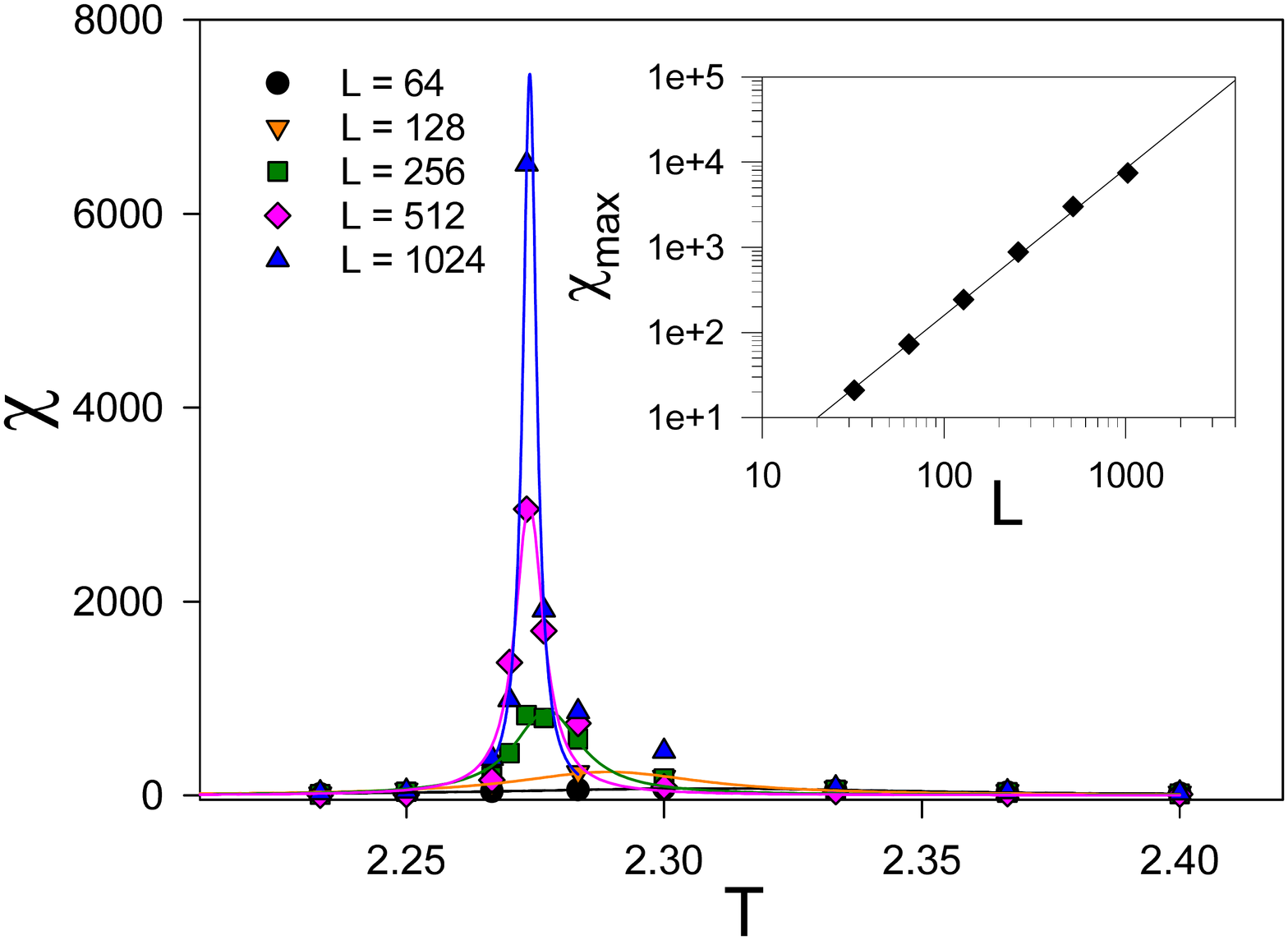}
\caption{Magnetic susceptibility $\chi$ for the zero magnetic field 2-D Ising model as a function of the temperature for different lattice sizes. The continuous lines correspond to a Lorentzian fitting close to the maximal. The inset shows a log-log plot of the susceptibility maximum $\chi_{max}$ as a function of the system size (error bars are smaller than the symbol size); the linear fitting ($r^2=0.9978$) provides an estimate of the critical exponent $\gamma/\nu= 1.72 \pm 0.05$. }
\label{FigSus}
\end{figure}

The results are compared with the exact result\cite{Huang} in Fig. \ref{FigMagnet}. Although the qualitative agreement is evident, the calculation of a quantity more sensitive to the fluctuations (such as a critical exponent)  is needed in order to ensure that the random numbers implementation does not introduce a statistical bias. Hence, we also computed the magnetic susceptibility (cf. Eq. \ref{EquSus}), which is displayed in 
Fig. \ref{FigSus}. The maximum of the susceptibility it is known to scale as $\chi_{max}\sim L^{\gamma/\nu}$ with the system size, where $\gamma$ and $\nu$ are the critical exponents of the susceptibility and the correlation length respectively \cite{Landau}. 

In order to estimate $\chi_{max}(L)$ we performed a Lorentzian fitting of the susceptibility curves for each value of $L$ (see Fig.\ref{FigSus}). The log-log plot of $\chi_{max}(L)$ vs. $L$ shown in the inset of Fig. \ref{FigSus} displays the expected power law behavior; a linear fitting of this plot provides an estimate $\gamma/\nu= 1.72 \pm 0.05$, in a very good agreement with the exact result $\gamma/\nu=1.75$ \cite{Reichl}.

\begin{figure}[h]
\centering
\includegraphics[width=0.5\textwidth,angle=-90]{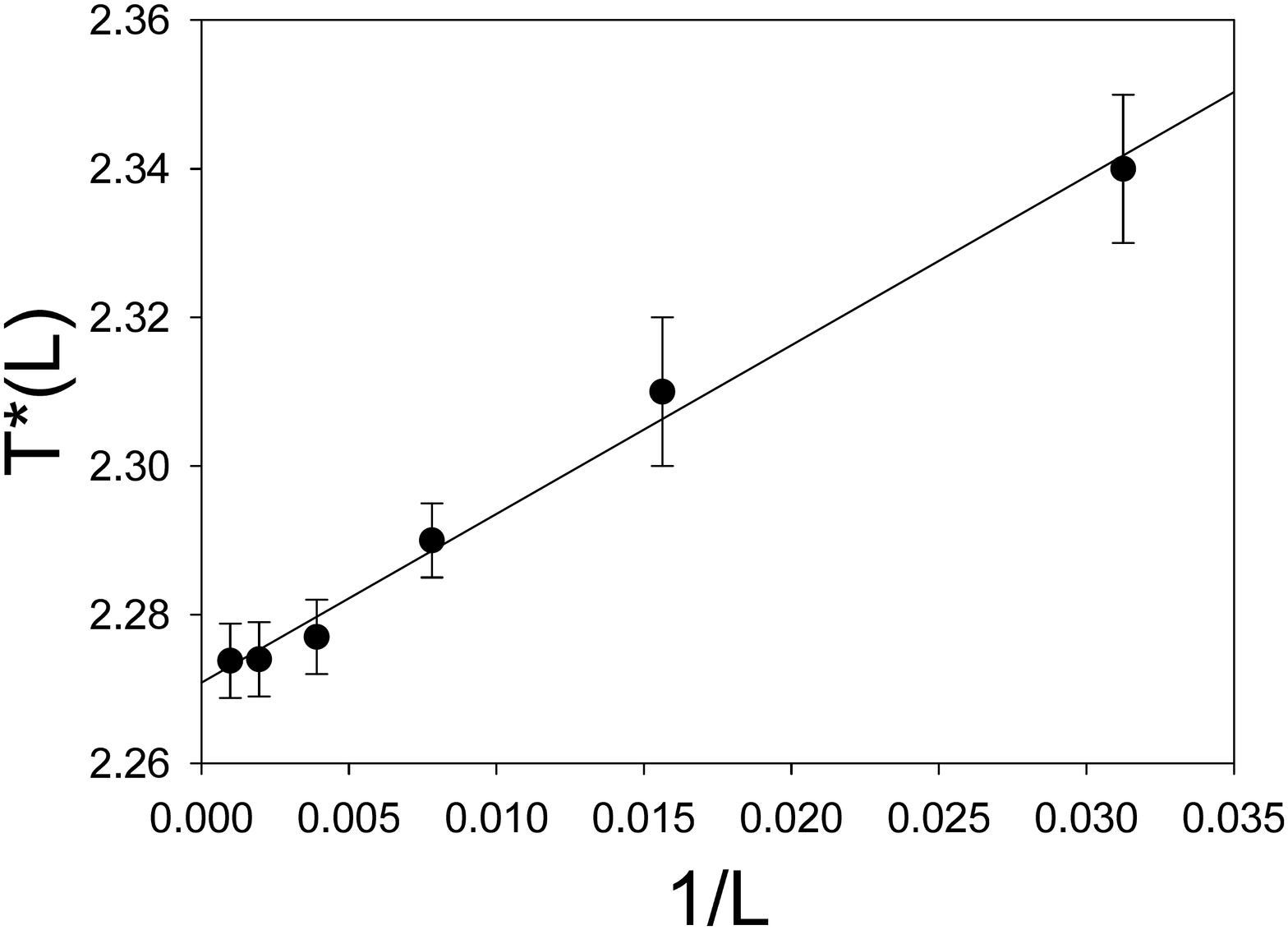}
\caption{Pseudo critical temperature $T^*(L)$ as a function of $1/L$. The linear fitting ($r^2=0.99$) provides the estimate $T_c=2.27 \pm 0.1$ (see text for details). }
\label{Tc}
\end{figure}

Finite size scaling provides also a method to estimate the critical temperature $T_c$. 
The maximum of the susceptibility is located at a pseudo critical temperature $T^*$ that depends on the system size and converges to the critical temperature $T_c$ as   $T^*(L) \sim T_c + b/L$, for large enough values of $L$, with $b>0$ \cite{Landau}. From the Lorentzian fittings we also estimated $T^*(L)$. In Fig.\ref{Tc} we plot $T*(L)$ vs. $1/L$. A linear extrapolation to $1/L \to 0$ allowed us  the estimate $T_c=2.27 \pm 0.1$, in excellent agreement with the exact result \cite{Huang} $T_c=2/\ln(1+\sqrt{2}) = 2.269...$. 
This set of results confirms the correct implementation of the algorithm.

Further, we have also compared the number of spins updates in a microsecond obtained from the current implementation to previously obtained values from \cite{Lin2013} regarding CPU, GPU and previous FPGA implementations.The results are shown in table \ref{TableRes}, where all values except those in the last row were extracted from the work of Lin et al. \cite{Lin2013}. 
To put the results shown in table \ref{TableRes} in context, we give next some details of the hardware and methods used for their obtention: the CPU platform results were obtained using an Intel core i5 at 2.67 GHz (CPU)  applying some optimization techniques like a linear congruential generator (LCG), sequential updating and cashed Boltzmann. "Previous FPGA" values were obtained using a DK-DEV-3CI20N board from Altera,  while results in relationship to the "64 GPUs" system  were achieved using  64 interconnected GPU processors in a supercomputer. For a discussion about the advantages/disadvantages of the current available hardware computing devices, in relationship to magnetic systems and in a more general context, we refer to the works of \cite{Ferrero1,Lin2013,Jones2010}.

\begin{table}[h]
\centering
\caption{Number of spins updated per microsecond for the 1024x1024 Ising model.}
\label{TableRes}
{\begin{tabular}{ | c | c | c |}
\hline
Platform  & \# updated spins & Ratio \\\hline
\hline
CPU 			&  62		& 1    \\
Single GPU 		& 7977		& 129  \\
Previous FPGA 	& 94127 	& 1518 \\
64 GPUs 		& 206000	& 3322 \\
Our FPGA		& 614400	& 9909 \\
\hline
\hline
\end{tabular}}{}
\end{table}

In relationship to the computation time required to execute the numerical simulations presented in this work, we have estimated the times employed to compute one MCS and a whole simulation for obtaining one data value at a given temperature. The results are shown in table \ref{TableTiempos}, as a function of the system lattice size.

\begin{table}[h]
\centering
\caption{Time employed to compute one MCS and a whole simulation carried to obtain data at a given temperature, as a function of system lattice size.}
\label{TableTiempos}
{\begin{tabular}{ | c | c | c | c | c | c | c | c |}
\hline
Size Lattice    & 16  & 32  & 64  & 128	 & 256   & 512   & 1024	\\\hline
\hline
1 MCS (ns)	& 6.6 & 6.6 & 6.6 & 26.6 & 106.6 & 426.6 & 1706.6\\
TOTAL (ms)	& 0.6 & 0.6 & 0.6 & 2.7  &  10.8 & 43.1  & 172.4 \\
\hline
\end{tabular}}{}

\end{table}

\section{Possible extensions}
The implementation described in the sections above can be extended with relatively modest modifications to systems of current research interest. Here we will describe three basic extensions that will illustrate the required changes in the implementation.

Perhaps the simplest extension involves the inclusion of next-nearest-neighbor interactions.  The next-nearest neighbors of the spin at position $(i,j)$ are defined as the four spins at positions $(i\pm 1,j\pm 1)$, that is along the diagonals of the spin at $(i,j)$. Of particular interest is the case in which the interaction with first neighbors is ferromagnetic and anti-ferromagnetic with next-nearest neighbors. This can be modelled by introducing a negative exchange constant that mediates the interaction of each spin with the second neighbors, as represented in the following Hamiltonian:

\begin{equation}
H=-J_1\,\sum_{\langle i,j \rangle} S_i S_j-J_2\,\sum_{\langle\langle i,j \rangle\rangle} S_i S_j \, ,
\end{equation}

\noindent where $J_1 > 0$, $J_2 < 0$, and $\langle\langle \ldots \rangle\rangle$ denotes that the sum runs over all next-nearest neighbor pairs. This model represents a binary spin version of the $J_1/J_2$ model \cite{Shang,Guerrero}.

The implementation of the binary $J_1/J_2$ model requires a small number of changes in the logic of the FPGA circuit shown in Fig.~\ref{FigIsing}. The main differences are the following: 
\begin{enumerate}[a)] 
\item  The values of the four next-nearest neighbor of the spin to be updated (L',T',R', and B') need to be considered in the circuit.
\item  Instead of the checkboard division into two sublattices the system can be divided into four sublattices in such a way that spins belonging to the same sublattice are neither nearest neighbors nor next-nearest neighbors---hence, requiring four memory modules to store the spin values of each sublattice.
\item The change in energy associated with a single spin flip is equal to $\Delta E=2 \epsilon'$, where  $\epsilon'$ is given by the following expression,
\begin{equation}
\epsilon'=\left(J_1(S_1+S_2+S_3+S_4)\right.+J_2\left.(S_1'+S_2'+S_3'+S_4') \right) \cdot S_0 \,.
\end{equation}
\item The look-up table for the energy values needs to be expanded since now the change in energy can take a wider range of values. In practical implementation it is usual to take $J_1$ or $J_2$ equal to 1 and then allow the other exchange constant to adjust the relative strength of the ferromagnetic/anti-ferromagnetic interactions.
\end{enumerate}

A second example of possible extensions is the implementation of the Potts model  \cite{Wu1982}.
The Potts model  is a generalization of the Ising model in which each individual spin can take an integer value $S_i=1,\ldots, q$. Its Hamiltonian is given by the following:
\begin{equation}
H=-J\sum_{\langle i,j \rangle} \delta(S_i,S_j) \,,
\end{equation}
\noindent where $\delta(S_i,S_j)$ is the Kronecker delta, whose value is 1 if $S_i=S_j$ and zero otherwise. The implementation of the Potts model implies some more significant changes to the logic in Fig.~\ref{FigIsing}. The main ones are the following: 

\begin{enumerate}[a)] 
\item   Instead of using Boolean variables as in the case of the Ising model, the Potts model requires that each spin can take one of $q$ different states
\item For the update of spin $S_i$ into the new state $S_i'$ the Monte Carlo algorithm proceeds by choosing at random one of the $q-1$ states for which $S_i \neq S_i'$, then computing the corresponding energy difference $\Delta E$ that the change implies, and finally the decision to accept it or not follows the same logic as the one for the Ising model.
\end{enumerate}

 One final example is the 3-D Ising model. Its implementation requires a small modification of the spin block in comparison with the 2-D Ising, as 6 neighboring spins should be considered instead of 4.  This change will affect the size of the lookup table because the energy range will be larger.  Memory management for this model will be also more complex and as in principle memory resources may not be enough to handle the whole system divided in two sub-lattices, but a layered scheme should be used where the whole system is analyzed as composed of M  2-D Ising models.

\section{Conclusions and discussion}
The results of this work confirm the potential of FPGA boards for the simulation of statistical mechanics models, demonstrating a significant improvement in terms of the numbers of spins updated per second was obtained in comparison to previous work (Table \ref{TableRes}). The improvements in performance are a consequence of efficient spin implementation and resource utilization. The spin representation was clearly described and analyzed in terms of the logic resources involved (cf. Figs \ref{FigIsing} and \ref{FigMemoria}, and Table \ref{tableSummary}), and  the use of a  combined global-local random number LFSR generator proved to be a significant factor in increased performance. The procedure based on the combination of two LFSR (one local and one global with the respect to the implementation of a block of spins)  was used, rigorously tested in terms of the quality of the random numbers produced using a  standard suite of  statistical tests \cite{WebNIST}. In relation to the previous work by Lin et al. \cite{Lin2013}, it is worth noting that their analysis showed that using a true RNG does not improved much the quality of the random numbers obtained. On the other hand, our approach used a local LFSR instead of a CA, and this fact permitted us to utilize efficiently the board resources to get faster spin updates. 

These improvements summed to the parallelism provided by the FPGA implementation offer a significant potential for extensions to other models of magnetic systems. The 2-D Ising model requires a minimal number of modifications to be extended to more complicated models like the binary spin version of the $J_1/J_2$ model, the Potts model, or the 3-D Ising model.  Further work is required to exploit the massive parallelism of FPGAs in  systems with more complex (long-range) interactions like the Ising dipolar  model \cite{Cannas2008}, which would require a complete redesign of the logic involved.


\section*{Acknowledgements}
The authors gratefully acknowledge the constructive comments made by the anonymous reviewers.
The authors acknowledge support from Junta de Andalucia through grants P10-TIC-5770, from MICIIN(Spain) through grants TIN2010-16556  and  TIN2014-58516-c2-1-R (all including FEDER funds), and from CONICET (Argentina) through grant PIP11220110100213.

The final version of this work has been published \cite{ortegafpga}

\bibliographystyle{IEEEtran}

\end{document}